\documentclass[a4paper]{jpconf}
\usepackage{graphicx}
\usepackage{iwsmihead}
\begin{document}

\title{Gene-network inference by message passing}

\author{A Braunstein$^{1,2}$, A Pagnani$^1$, M Weigt$^1$ and R
Zecchina$^{1,2}$}

\address{$^1$ Institute for Scientific Interchange, Viale S. Severo
65, I-10133 Torino, Italy} 

\address{$^2$ Politecnico di Torino, Corso Duca degli Abruzzi 24,
I-10129 Torino, Italy}

\ead{weigt@isi.it}

\begin{abstract}
The inference of gene-regulatory processes from gene-expression data
belongs to the major challenges of computational systems biology. Here
we address the problem from a statistical-physics perspective and
develop a message-passing algorithm which is able to infer sparse,
directed and combinatorial regulatory mechanisms. Using the replica
technique, the algorithmic performance can be characterized
analytically for artificially generated data. The algorithm is applied
to genome-wide expression data of baker's yeast under various
environmental conditions. We find clear cases of combinatorial
control, and enrichment in common functional annotations of regulated
genes and their regulators.
\end{abstract}

\section{Introduction}

\begin{figure}[h!]
\begin{center}
\includegraphics[width=9cm]{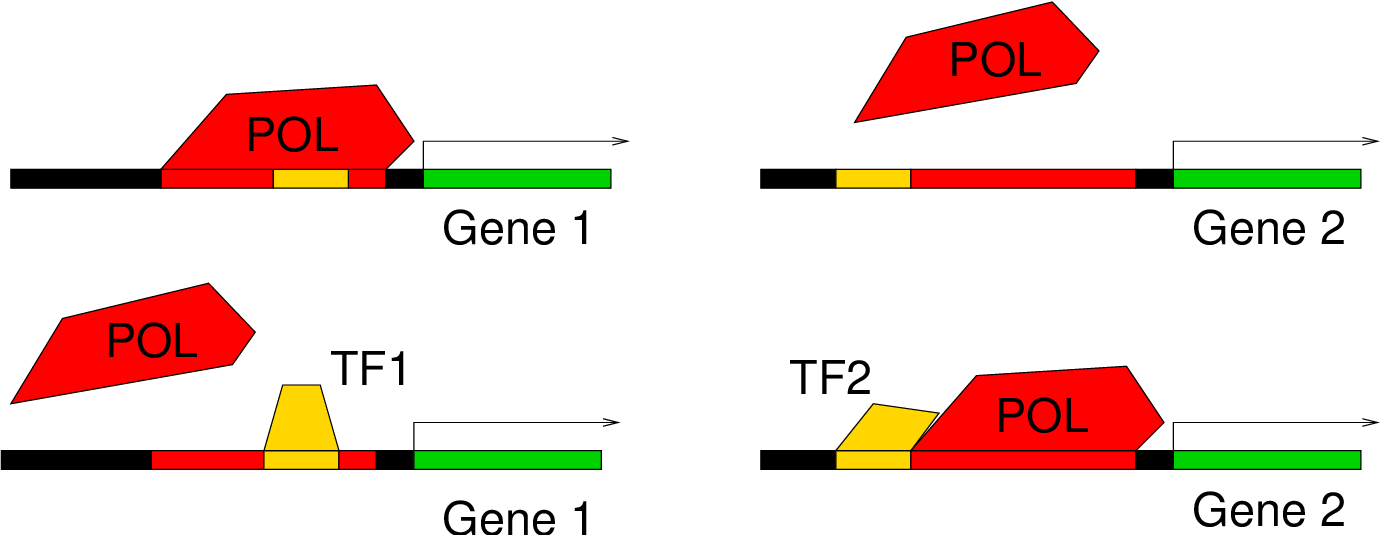}
\end{center}
\caption{\label{fig:reg} Schematic representation of transcriptional
repression (left) and activation (right) of a gene by a single
transcription factor (TF). The RNA polymerase is denoted by POL.}
\end{figure}

Transcriptional gene regulation is at the basis of cell development
and differentiation \cite{Alberts}. It constitutes the important
feed-back mechanism from the level of proteins to the transcription of
genes to mRNA, and it allows to reach differentiated gene expression
patterns starting from identical genetic information. The simplest
mechanisms in this context are transcriptional repression and
activation of a gene by a single transcription factor (TF) which are
schematically depicted in figure~\ref{fig:reg}:
\begin{itemize}
\item In the case of a {\it repressor}, the RNA polymerase (i.e. the
molecular machine transcribing genes to mRNA, abbreviated as POL in
the figure) and a transcription factor (i.e. a regulatory protein)
have overlapping binding sites on the DNA. Due to steric exclusion
effects the presence of the TF hinders the polymerase to bind to DNA
and thus to transcribe the gene. Elevated levels of TF concentration
thus lead to a repression of the transcription rate of the considered
gene.
\item In the case of an {\it activator}, polymerase and TF bind
cooperatively to the DNA. Their binding sites are close, and both
proteins show some attractive short-range interaction of a few $kT$.
A high TF concentration thus increases the binding probability of
polymerase, and transcription is enhanced.
\end{itemize}

Also the expression of transcription factors is regulated by other
transcription factors. The set of all genetic interactions is called
the {\it gene-regulatory network} (GRN). It is considered to be a
prototypical example of a complex system. GRN are characterized in
particular by the following properties:
\begin{itemize}
\item They are {\it sparse}: Each gene is controlled only by a limited
number of other genes, which is very small compared to the total
number of genes present in an organism.
\item Gene networks are {\it directed}: Regulatory control is
obviously directed from the regulators to the regulated genes.
\item Genes are regulated via {\it combinatorial control} mechanisms:
The expression level of a gene frequently depends on the joint
activity of various regulatory proteins. The best-known example is
the Lac operon which is activated only if glucose is absent AND 
lactose is present.
\end{itemize}

Understanding GRN is a major task in modern biology, but its
experimental determination is extremely complicated. Genome-scale
networks are only known for {\it E. coli} \cite{Alon1} and baker's
yeast \cite{Kepes,Alon2}, whereas for higher organisms only few
functional modules are reconstructed, see {\it e.g.}
\cite{Davidson,Albert}. It is therefore tempting to ask in how far
gene-regulatory networks can be reconstructed starting from easily
accessible data -- in particular from genome wide expression data
measuring simultaneously the expression level of order $10^4$ RNAs
(micro-arrays). However, also this task is highly non-trivial and
limited by the following restrictions on the quality and quantity of
available data:
\begin{itemize}
\item The number $M$ of available expression patterns is in general
considerably smaller than the number $N$ of measured genes.
\item The available information is incomplete.  The expression of some
relevant genes may not be recorded, and external conditions
corresponding, {\it {\it e.g.}}, to nutrient or mineral conditions are
not given. Most importantly, micro-arrays measure the abundance of
mRNA, whereas gene regulation works via the binding of the
corresponding proteins (TFs) to the regulatory regions on the
DNA. Active protein and RNA concentrations are, however, not in a
simple one-to-one correspondence.
\item Data are noisy. This concerns biological noise due to the
stochastic nature of underlying molecular processes, as well as the
considerable experimental noise existing in current high-throughput
techniques.
\item Micro-arrays do not measure the expression profiles of single
cells, but of bunches of similar cells. This averaging procedure may
hide the precise character of the regulatory processes taking place in
the single cells.
\item Non-transcriptional control mechanisms (chromatin remodeling,
small RNAs etc.) cannot be taken into account in expression based
algorithms.
\end{itemize}                             
The listed points obviously lead to a limited predictability of even
the most sophisticated algorithms. State-of-the-art algorithms include
relevance networks measuring pairwise correlations \cite{RN1,RN2},
ARACNe which removes false positives in relevance
networks\cite{Aracne1,Aracne2}, Bayesian networks
\cite{Bayes1,Bayes2,Bayes3,Bayes4}, probabilistic Boolean networks
\cite{PBN1,PBN2}, and module networks \cite{Segal}. It is therefore
very important to test various algorithms on the basis of well
described data sets containing some or all of the before-mentioned
problems: Only a critical discussion on artificial data sets allows
for a sensible interpretation of the algorithmic outcomes when run on
biological data.

In the following, we first introduce a minimal functional model which
shall be used to fit the data. This model is incorporated in a scoring
function for networks, but the construction of high-scoring solutions
is a computationally hard task. In section \ref{sec:bp} we propose
therefore a message-passing approach to heuristically solve the
problem. In section \ref{sec:artificial}, the algorithm is analyzed on
artificial data, and finally it is applied to biological data of yeast
under various environmental conditions in
section \ref{sec:yeast}. Conclusions are drawn in the last section.

\section{The model}
\label{sec:model}

Before setting up a functional model we have to spend a few lines
discussing how data look like. Raw micro-array data require
normalization procedures, a standard way is filtering out genes having
very low variability, followed by log-ratio-normalizing the remaining
data. This means that data are divided by gene specific numbers
defining a reference level of expression for this gene, and then the
logarithm is taken. This guarantees that data of expression level
lower than the reference become negative, over-expression is shown by
positive values. Note also that the logarithmic transformation
regularizes the data distribution: The resulting single-gene
distributions look regular since the same fold-change in over- and
under-expression lead to symmetric log-normalized numbers. Let us
therefore assume that we have such log-ratio-values
\begin{equation}
x_i^\mu, \ \ \ \ \ \ i=0,...,N,\ \ \mu=1,...,M\ ,
\end{equation}
for $N+1$ genes measured in $M$ distinct micro-arrays.

The task is now to go from these data back to the interactions behind
it. To use a statistical-physics analogy, starting from snap shots of
the microscopic state of an Ising model we try to infer its
Hamiltonian. Note that due to the directed structure of gene networks
this task can be formally factorized over regulated genes: We can ask
first, which genes have a regulatory influence on gene 0, and how they
interact combinatorially. Then we ask the same question for the
regulators of gene 1, gene 2 etc., until we reach gene N. In the
following discussion we therefore concentrate without loss of
generality on one single regulated gene ($i=0$), and $N$ potential
input genes ($1\leq i \leq N$).

We further simplify the possible influence genes can have on the
target gene 0, we aim at a ternary classification of the influence of
a gene $i$ on 0:
\begin{equation}
J_{i\to 0} = \left\{
\begin{array}{rl}
-1 & {\mbox{ if gene $i$ represses the expression of gene 0,}}\\
0 & {\mbox{ if gene $i$ does not regulate gene 0,}}\\
1 & {\mbox{ if gene $i$ activates the expression of gene 0.}}
\end{array}
\right.
\end{equation}
As a minimal functional model, we assume that a gene is over-expressed
if the overall influence of its neighbors is beyond some threshold
$\tau$, and it is repressed if the overall influence is smaller then
$\tau$. We therefore expect
\begin{equation}
\label{eq:constraints}
x_0^\mu > 0 \ \ \ \ \ \ \ \leftrightarrow \ \ \ \ \ \ \ 
\sum_{i=1}^N J_{i\to 0} x_i^\mu > \tau
\end{equation}
to hold for as many expression patterns $\mu=1,...,M$ as possible. In
this sense, each pattern gives a {\it constraint} on the coupling
vector $\vec J\ =(J_{1\to 0},...,J_{N\to 0})$, and the problem of
finding a good candidate vector $\vec J$ can be understood as an
instance of a {\it constraint satisfaction problem}. Also the
threshold can be inferred similarly to $\vec J$, but to avoid heavy
notation we set it to zero in the following.

A cost function
for this problem counts the number of errors made in
(\ref{eq:constraints}),
\begin{equation}
{\cal H}(\vec J\ ) = \sum_{\mu=1}^M \Theta\left(
-x_0^\mu\ \sum_{i=1}^N J_{i\to 0} x_i^\mu 
\right) \ ,
\end{equation}
with $\Theta$ being the step function. Obviously, this ternary
classification is over-simplified in the sense that no weak or strong
repressors and activators are considered, and more complex functions
like the XOR of the inputs (or an continuously-valued generalization 
of it) cannot be represented in this way. However, due to the
before-mentioned problems with data quality and quantity we have to
restrict ourselves to a not too complex class of models in order to
avoid over-fitting.

As already mentioned, GRN are sparse, i.e., only a few of the $i\in
\{1,...,N\}$ will have a non-vanishing coupling to gene 0. We
therefore have to control also the effective coupling number
\begin{equation}
N_{eff}(\vec J\ ) = \sum_{i=1}^N |J_{i\to 0}|\ ,
\end{equation}
which counts only the number of non-zero entries in $\vec J$. An {\it
a priori} bias towards diluted graphs is also reasonable from the
machine-learning standpoint. The restriction of the entropy of the
search space lowers the probability of over-fitting.

\section{Inference by belief propagation}
\label{sec:bp}

The inference task is now to characterize the properties of vectors
$\vec J$ which are both low-cost and diluted. To achieve this we
introduce a weight
\begin{equation}
\label{eq:j-distribution}
W(\vec J\ ) = \exp\left\{ - \beta {\cal H}( \vec J\ ) -h N_{eff}(\vec
J\ ) \right\}\ ,
\end{equation}
which still depends on two external parameters $\beta$ and $h$ which
act as a formal inverse temperature and a diluting field. The size of
these parameters determines the relative importance of low-cost vectors
compared to sparse ones.

In order to get information about the statistical properties of the
single-gene couplings $J_{i\to 0}$ we have to calculate marginals
\begin{equation}
\label{eq:marginal}
P_i(J_{i\to 0}) 
\propto \sum_{\{J_{j\to 0}\in\{0,\pm 1\};j\neq i\}} W(\vec J)\ .
\end{equation}
The probability $P_i(J_{i\to 0}\neq 0) = 1-P_i(0)$ of having a
non-zero coupling can be used to rank genes according to their
relevance for gene 0. However, the direct calculation requires a sum
over $3^{N-1}$ configurations and is therefore infeasible even for
relatively small systems. The construction of high-weight vectors
itself is already an NP-hard task, so we need to apply heuristic
methods to approximate $P_i$.

The main idea is to use belief propagation (BP). Variables and
constraints exchange messages,
\begin{eqnarray}
\label{eq:bp1}
P_{i\to \mu} (J_{i\to 0}) & \propto & e^{-h |J_{i\to 0}|}\ 
\prod_{\nu\neq \mu} \rho_{\nu\to i} (J_{i\to 0})\ , \nonumber\\
\rho_{\mu\to i} (J_{i\to 0}) & \propto & 
\sum_{\left\{ J_{j\to 0},\ j\neq i\right\} } \exp\left\{ -\beta
\Theta\left[ -x_0^\mu\cdot \sum_k J_{k\to 0} x_k^\mu \right]\right\} 
\prod_{j\neq i}P_{j\to \mu}(J_{j\to 0})\ ,
\end{eqnarray}
which have to be determined self-consistently. They can be used to
calculate the BP approximation for the marginal distributions,
\begin{equation}
P_{i} (J_{i\to 0}) \propto e^{-h |J_{i\to 0}|}\ 
\prod_{\mu}\rho_{\mu\to i}(J_{i\to 0})\ .
\end{equation}

Looking a bit closer to the second of equations \ref{eq:bp1}, we see that
it still contains the exponential summation over coupling
configurations. Due to the factorization of messages used in BP it
results in an average over independent random variables. The quantity to
be averaged depends only on the sum over these variables, so we may use a
Gaussian approximation
\begin{equation}
  \label{eq:bp_gauss}
\rho_{\mu\to i} (J_{i\to 0})  \propto \int_{-\infty}^\infty \frac{dh}{\sqrt{2\pi}
\Delta_{\mu\to i}} \exp\left\{ -\frac{(h-h_{\mu\to i})^2}{2\Delta_{\mu\to i}^2} 
-\beta \Theta\left[ -x_0^\mu\cdot (J_{i\to 0} x_i^\mu + h )\right]\right\} 
\end{equation}
with 
\begin{eqnarray}
h_{\mu\to i} &=& \sum_{j\neq i} x_j^\mu\ \langle J_{j\to 0} \rangle_{j\to\mu} \ ,
\nonumber\\
\Delta_{\mu\to i}^2 &=& \sum_{j\neq i}\ \left(\ \langle J_{j\to 0}^2 
\rangle_{j\to\mu} -  \langle J_{j\to 0} \rangle_{j\to\mu}^2\ \right) \ ,
\end{eqnarray}
which brings the computational cost for a single message update down
to linear time in $N$. Very similar constructions were used in
\cite{CDMA,UdKa,BrZe}.

Having calculated the marginal probabilities, we can also determine
the energy of the average coupling vector
\begin{equation}
E = \sum_{\mu=1}^M \Theta \left( -x_0^\mu\cdot 
\sum_{i=1}^N \langle J_{i\to 0} \rangle_i\ x_i^m \right)
\end{equation}
and the Bethe entropy
\begin{equation}
\label{eq:entropy}
S = \sum_{\mu=1}^M S_\mu - (N-1)\sum_{i=1}^N S_i
\end{equation}
characterizing the number of ``good'' coupling vectors. In the last
expression, the site entropy $S_i$ is given by
\begin{equation}
\label{eq:S_i}
S_i = \sum_{J_i=-1,0,1} P_i(J_{i\to 0})\ \ln P_i(J_{i\to 0}) \ ,
\end{equation}
and the pattern entropy
\begin{eqnarray}
\label{eq:S_mu}
S_\mu &=& \sum_{\vec J} P_\mu(\vec J\; )\ \ln P_\mu(\vec J\; )\ , \\
P_\mu(\vec J\; ) &=& \exp\left\{ -\beta
\Theta\left[ -x_0^\mu\cdot \sum_k J_{k\to 0} x_k^\mu \right]\right\} 
\prod_{i}P_{i\to \mu}(J_{i\to 0})\ ,
\nonumber
\end{eqnarray}
can be calculated in analogy to $\rho_{\mu\to i}$ via a Gaussian
approximation of the sum over $\vec J$.

These quantities can be used to fix the free parameters. Imagine, {\it
e.g.}, that we want to achieve some dilution $N_{eff}(\vec J)$. Then
we can start at high temperature $\beta^{-1}$ and low diluting field
$h$, and during the iterative solution of the BP equations we adopt
the parameters slowly such that at the end $N_{eff}(\vec J)$ takes the
desired value, and the entropy $S$ vanishes. This end point
corresponds to the ground state at given dilution. In the case that
the noise level in the data is known, we can also use this to fix a
(possibly pattern- and gene-dependent) temperature right from the
beginning.

\section{Artificial data}
\label{sec:artificial}

Before applying the algorithm to real biological data, it is useful to
check its performance on artificial data. The basic idea is to first
generate data by some known network, and then to feed it to our
algorithm. The inferred couplings can be compared to the known ones of
the generator.

Here we consider a very simple data generator which has the advantage
of being analytically tractable with tools from the statistical
physics of disordered systems, more precisely with the replica trick.
To do so, we still simplify a bit the situation and look only to
binary data $x_i^\mu=\pm 1$, and to $N$ inputs ($i=1,...,N$) and one
output ($i=0$) which are related to each other by the relation
\begin{equation}
\label{eq:generator}
x_0^\mu = {\rm sign} \left( \sum_{i=1}^N J^0_{i\to 0}\ x_i^\mu + \eta^\mu
\right)\ ,
\end{equation}
for all $\mu=1,..,M$. The inputs are assumed to be independent and
unbiased random numbers. To render the inference task non-trivial, we
assume two major differences with respect to the ternary
classification done by BP:
\begin{itemize}
\item {\it Heterogeneity of couplings}: The couplings $J^0_{i\to 0}$
may take values different from $0,\pm 1$, allowing for weak and strong
influences of repressor and activator genes. In this study we use
\begin{equation}
\rho(J^0_{i\to 0}) = (1-k_1-k_2) \delta(J^0_{i\to 0}) + \frac{k_1}2
[\delta(J^0_{i\to 0}+1)+\delta(J^0_{i\to 0}-1)] + \frac{k_2}2
[\delta(J^0_{i\to 0}+2)+\delta(J^0_{i\to 0}-2)].
\end{equation}
Generalizations are straight-forward. To meet the biological constraint
of diluted interactions we assume $k_{1,2}\ll 1$.
\item {\it Noise}: Biological data are noisy. We therefore include
white Gaussian noise with
\begin{eqnarray}
\overline {\eta^\mu} &=& 0 \ ,\nonumber\\
\overline {\eta^\mu \eta^\nu} &=& \gamma\ N\ \delta_{\mu,\nu}\ .
\end{eqnarray}
The scaling of the variance with $N$ ensures a finite signal-to-noise
ratio. For the special case $\gamma = k_1+ 4 k_2$, the statistical
properties of signal and noise are identical.
\end{itemize}
Obviously, biological data coming out of a full network are correlated
even at the input level to gene 0, but we do not consider such
correlations here. Since original and inferred couplings take different
values, we use the following standard notation for comparing both
vectors:
\begin{equation}
\begin{array}{lll}
J_{i\to 0}^0 = 0\ \ \ \  & J_{i\to 0} = 0\ \ \ \  & 
{\mbox{ true negative (TN)}}\\
J_{i\to 0}^0 \neq 0 & J_{i\to 0} = 0 & {\mbox{ false negative (FN)}}\\
J_{i\to 0}^0 = 0 & J_{i\to 0} \neq 0 & {\mbox{ false positive (FP)}}\\
J_{i\to 0}^0 \neq 0 & J_{i\to 0} \neq 0 & {\mbox{ true positive (TP)}}
\end{array}
\end{equation}
A major aim in network inference is to predict a {\it fraction} of
all couplings with high precision, i.e. to have an as high as possible
number of TP with a low number of FP. The quality measure we use will
be the confrontation of the {\it recall}, or {\it sensitivity},
\begin{equation}
RC = \frac{N_{TP}}{N_{TP}+N_{FN}}\ ,
\end{equation}
and of the {\it precision}, or {\it specificity},
\begin{equation}
PR = \frac{N_{TP}}{N_{TP}+N_{FP}}\ .
\end{equation}
The recall describes the fraction of all existing non-zero couplings
which are predicted by the algorithm, whereas the precision tells us
which fraction of all predicted links is actually present in the data
generator. Both quantities are in competition: To have a very high
precision, only the strongest signals ($P_i(J_{i\to 0}=0)\ll 1$) should
be taken into account, but obviously this reduces the recall. A
high recall is achieved by accepting also weaker signals, and
obviously the FP rate will grow accordingly.

The nicest aspect of the simple data generator is that it can be
analyzed using the replica trick, at least in the thermodynamic limit
$N\to \infty$ with $\alpha=M/N = {\cal O}(1)$ and $k_{1,2}={\cal
O}(1)$. Details of this analysis are given elsewhere \cite{BrPaWeZe},
here we only show some results.

\begin{figure}[t!]
\begin{center}
	\includegraphics[width=0.8\columnwidth]{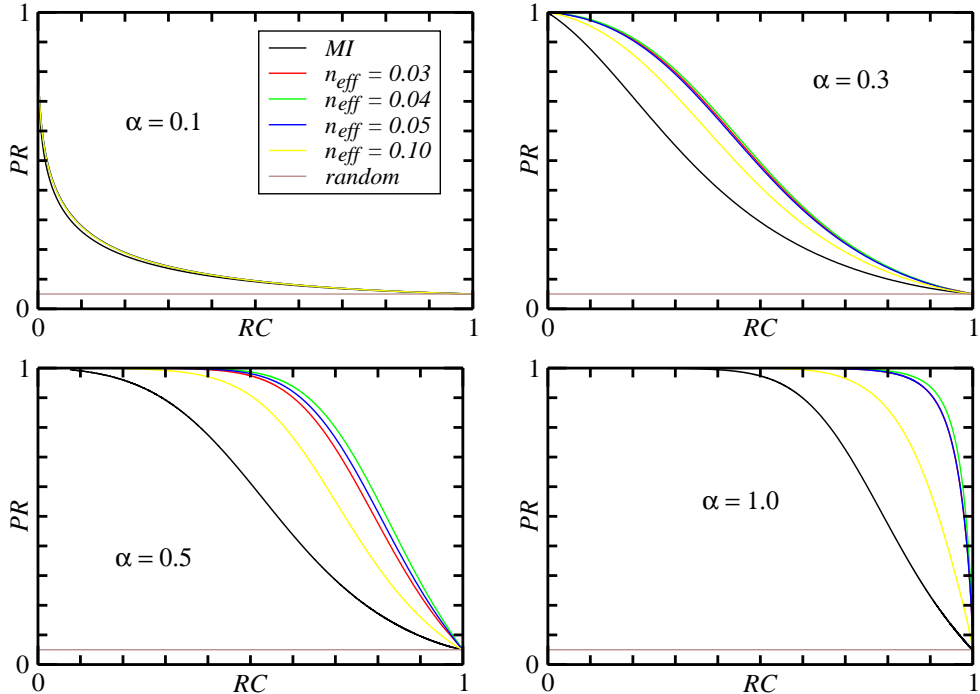}
\end{center}
\vspace{0.55cm}
\caption{Theoretical results of the performance of BP versus
pair-correlation based methods (MI=mutual information between single
input and output), for various values of $\alpha=M/N$. Other
parameters are $k_1=k_2=0.025$ and $\gamma = 0$.}
\label{fig:bp_vs_aracne}
\end{figure}

Figure \ref{fig:bp_vs_aracne} shows the performance of BP in the
noiseless case for $k_1=k_2=0.025$, i.e., only 5\% of the couplings in
the generator are non-zero. A perfect algorithm would reach recall one
with precision one, a completely random algorithm would have precision
5\% right from the beginning. Curves in between the two extrema see
some of the structure of the generator, but include also false
positives. We see that obviously a higher number of patterns improves
considerably the performance, but even for $\alpha=0.1$ curves start
at precision one. The performance depends on the dilution
$n_{eff}=N_{eff}/N$ of the inferred coupling, a dilution slightly
below the one of the generator turns out to be optimal. As a
comparison, we have also included the result of a simple pair-based
algorithm ranking input genes according to their mutual information
with the output, as done in some state-of-the-art approaches to
network inference \cite{RN1,RN2,Aracne1,Aracne2}. We see that BP
outperforms this approach, mainly because it is able to detect some
combinatorial effects which by definition are not seen by a
pair-correlation based method.

\begin{figure}[t!]
\vspace{1.5cm}
\begin{center}
\includegraphics[width=0.5\columnwidth]{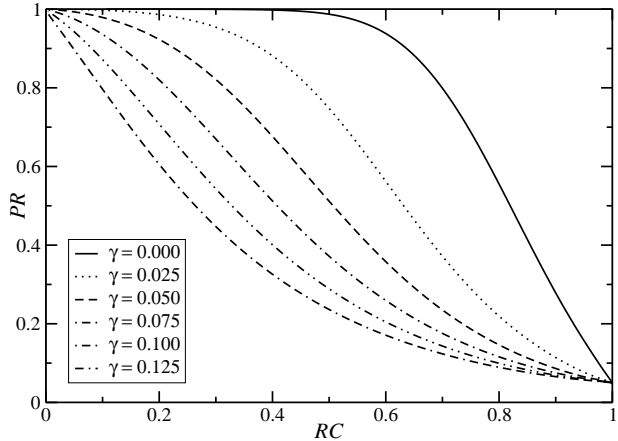}
\end{center}
\caption{Precision versus recall of BP for various noise strengths
$\gamma = 0.0,\ 0.025,\ 0.05,\ 0.075,\ 0.1,\ 0.125$ (curves from right
to left), ranging from no noise to equal signal and noise strengths.
Other parameters are $k_1=k_2=0.025$, $n_{eff}=0.04$ and $\alpha =
0.5$.}
\label{fig:with_noise}
\end{figure}

Figure \ref{fig:with_noise} evaluates the noise influence for the same
dilution parameters at $\alpha=0.5$. Curves go from zero noise to
$\gamma=k_1+4 k_2$ where input signal and noise have the same
variance. The performance of BP obviously goes down with increasing
$\gamma$, but the performance loss is continuous.

To resume this section, we see that BP is able to outperform simpler,
pair-based approaches. For very low pattern number $M$, however, this
performance gain is relatively small. Only for intermediate data
quantity the strength of BP becomes really relevant.

\section{The yeast network}
\label{sec:yeast}

After having analyzed the performance of BP in the case of artificial
data, now we discuss its application to real gene-expression
data. More specifically we look at genome wide data for baker's yeast
({\it Saccharomyces cerevisae}) using publicly available data of
Gasch, Spellman {\it et al.} \cite{Gasch}. Expression levels of 6152
genes have been recorded under 172 environmental conditions (like
temperature shock, osmotic pressure, starvation etc.).

First we have preprocessed data: We have eliminated genes which are
known to directly respond to external stress \cite{Gasch} since we are
interested only in internal regulatory mechanisms. Further on we have
filtered out all genes of small variance (we have used three-times the
minimum variance), and genes with more than 10 missing data points in
the 172 arrays. The resulting 2659 genes are used for inference,
whereas the smallest found variance was used as a (slightly
pessimistic) estimate of the noise level. We used it to fix the
temperature, so only the diluting field remains as a free parameter.

To increase $\alpha$, we did not use all filtered genes as possible
input variables, but restricted the search to 460 potential regulators
(known TFs, signaling proteins, proteins structurally similar to known
regulators etc.) \cite{Segal}. Further on, to estimate the quality of
our prediction, for each of the 2659 output genes we divided the data
set ten times randomly into a training set of 142 patterns, and a test
set of 30 patterns. BP was run using only the training patterns, and
the average coupling vector was used to estimate the generalization
error on the test data. Doing so, we had to run BP 26590 times, which
took less than two weeks on a single desktop PC.

As an output, we found on average 2.3 regulators per gene. The
predictability on the test data was about 76\%, which is only slightly
better than the one obtained using as a predictor the three most
correlated genes (74\% correct predictions on average). This result
has, however, to be compared to an error bound obtained on the basis
of the noise estimate: In particular genes of expression close to zero
have a considerable probability of being measured with opposite sign
in the micro-array, leading thus to an {\it a priori} error in the
cost function. Taking the specific values of yeast, we expect about
20\% of all measured expression levels to have the wrong sign compared
to the actual mRNA abundance, so the mean predictability cannot grow
beyond 80\%.

\begin{figure}[htb]
\vspace{1.5cm}
\begin{center}
\includegraphics[width=0.6\columnwidth]{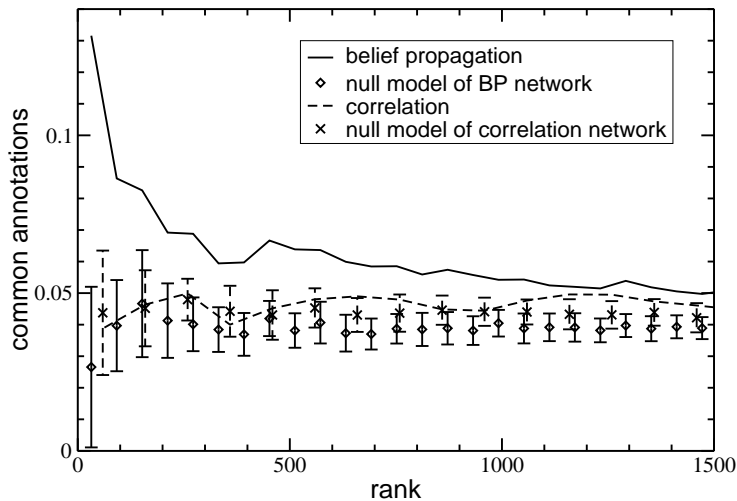}
\end{center}
\caption{Fraction of links with common functional annotation, as a
function of genes ranked according to their prediction error.}
\label{fig:annotation}
\end{figure}

Is there some biological signal in the inferred couplings? To answer
this question, we have looked to all those links which have
gene-ontology annotations (see http://www.geneontology.org/) for both
extremities of the link, and we have determined the fraction of links
with at least one common annotation. In figure \ref{fig:annotation} the
results are presented, with the genes ranked according to their
predictability (highest rank = smallest prediction error). As a
comparison the results for a null model (randomly rewired graphs having
the same in- and out-degrees as the inferred one) are depicted. Symbols 
and bars represent averages and variances of the null model. It is
obvious that the BP results show a significant enrichment in common
functional annotations, with a strong correlation to the quality of
the prediction error. No such signal is detected if we use the three
most correlated input genes as predictors.

\begin{figure}[htb]
\vspace{1.5cm}
\begin{center}
\includegraphics[width=0.7\columnwidth]{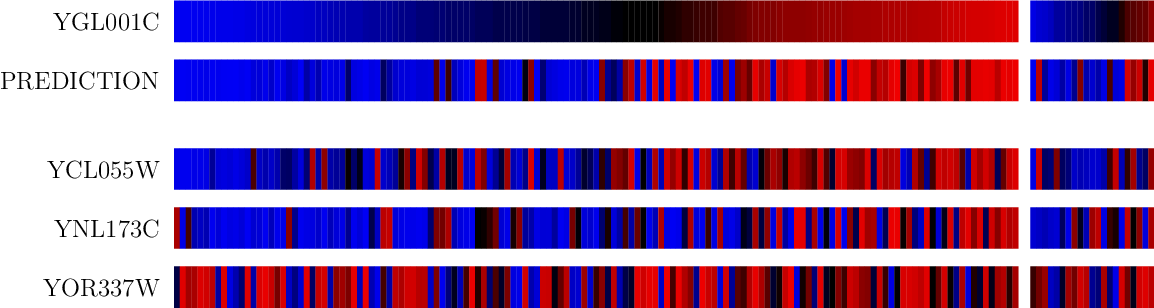}
\end{center}
\caption{Measured vs. predicted gene expression with belief
propagation: The first bar is the measured gene, divided in training
and test set. The second bar is the prediction using BP, the following 
three rows are the most relevant input genes according to BP. Repression is
denoted in blue, activation in red. Patterns are ordered according to the
expression level of the output gene.}
\label{fig:comb_cont_bp}
\end{figure}

\begin{figure}[htb]
\vspace{1.5cm}
\begin{center}
\includegraphics[width=0.7\columnwidth]{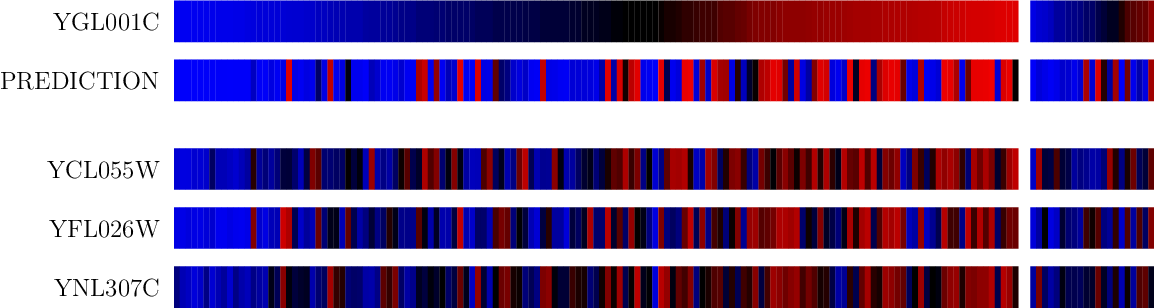}
\end{center}
\caption{Measured vs. predicted gene expression by the most correlated
genes: The first bar is the measured gene, divided in training and
test set, the second bar is the prediction by the three most
correlated genes, the following three bars are the input genes of
highest correlation to the output in the training set.}
\label{fig:comb_cont_corr}
\end{figure}

Last but not least, figures \ref{fig:comb_cont_bp} and
\ref{fig:comb_cont_corr} show an example of combinatorial control. The
upper part of figure \ref{fig:comb_cont_bp} shows the expression of
gene YGL001 together with its prediction by the BP selected input
regulators, the left bars depicts the training, the right ones the test
set. The lower bars are the three most important input genes according
to BP. Note that some of the cases where the first input gene has an
opposite sign compared to the output gene are cured by the two other
genes, which gives a clear illustration of combinatorial control. 
Figure \ref{fig:comb_cont_corr} shows the analogous result for
the three most correlated genes. The first input here is identical to
the one found by BP, but its errors are not corrected by the other
inputs. In fact we find the prediction on the test set to be much
better for BP than for pair correlations.

\section{Conclusion and outlook}
\label{sec:conclusion}

In this article, we have introduced, analyzed and applied a
message-passing algorithm for the inference of gene interactions from
genome-wide expression experiments. With the aim to infer a sparse,
directed network showing combinatorial control we have set up a
minimal model classifying regulators into three classes: repressors,
activators or non-regulators. Even if simplified compared to
biological reality, this simple model is expected to reflect part of
the relevant biological processes.

In the case of artificial data we have shown that the algorithm
predicts some true positive links even in the case of very few input
patterns. The performance increases if more patterns are present. We
have also seen that BP outperforms simpler pair-correlation based
approaches. The advantage of BP was small for few data, and big for an
intermediate number of data. In the limit of infinitely many data both
algorithms are expected to perform well.

In the case of biological data, more precisely yeast expression
profiles, a small but systematic advantage of BP compared to
correlations was observed. This suggests that we are still in the
low-data regime, and more data would be needed to profit from the high
potential of message passing techniques. However, we have seen that
genes resulting in a smaller than average prediction error show
strongly enriched common functional annotations between regulators and
regulated genes. We therefore expect that more and less noisy data
lead to a more pronounced difference between BP and traditional 
pair-correlation based tools.

One of the weak points in the derivation of the algorithm is that we
neglected correlations between input genes. It is known that these
correlations are strong in biological data, so it appears to be
important to study systematically the influence of correlations on the
performance of BP. This will be done in a future work.

Further on, the fact that only few data are available requires the
integration of further biological knowledge. We already used a
precompiled list of potential regulators (instead of all genes), but it
could be highly interesting to include also sequence information on
putative binding sites.

\section*{Acknowledgments} 

We acknowledge interesting discussions with
Michele Leone, Francesca Tria and Yoshiyuki Kabashima.

\end{document}